\documentclass[twocolumn,secnumarabic,showpacs,amssymb, amsmath,tightenlines,aps,prb]{revtex4}
\usepackage{longtable}
\usepackage{bm}
\usepackage{epsfig}
\usepackage{graphicx}

\begin{document}

\title[Anomalous scaling behavior of the dynamical spin susceptibility of Ce$_{0.925}$La$_{0.075}$Ru$_{2}$Si$_{2}$]
{Anomalous scaling behavior of the dynamical spin susceptibility
of Ce$_{0.925}$La$_{0.075}$Ru$_{2}$Si$_{2}$}
\author{W. Knafo$^{1}$, S. Raymond$^{1}$, J. Flouquet$^{1}$, B. F{\aa}k$^{1}$, M.A. Adams$^{2}$, P. Haen$^{3}$, F. Lapierre$^{3}$, S. Yates$^{3}$, and P. Lejay$^{3}$}

\address{$^{1}$ CEA-Grenoble, DSM/DRFMC/SPSMS, 38054 Grenoble
Cedex 9, France \\$^{2}$ ISIS Facility, Rutherford Appleton
Laboratory, Chilton, Didcot,
 Oxon OX11 0QX, UK \\$^{3}$ CRTBT, CNRS, B.P. 166, 38042 Grenoble Cedex 9, France}

\date{\today}

\begin{abstract}

Inelastic neutron scattering measurements have been performed on
single crystals of the heavy fermion compound
Ce$_{0.925}$La$_{0.075}$Ru$_{2}$Si$_{2}$ in broad energy [0.1, 9.5
meV] and temperature [40 mK, 294 K] ranges in order to address the
question of scaling behavior of the dynamical spin susceptibility
at the quantum critical point of an itinerant magnetic system. For
two wavevectors $\mathbf{Q}$ corresponding to uncorrelated and
antiferromagnetically correlated spin fluctuations, it is found
that the dynamical spin susceptibility $\chi''(\mathbf{Q},E,T)$ is
independent of temperature below a cut-off temperature
$T_{\mathbf{Q}}$: the spin fluctuation amplitude saturates at low
temperatures contrarily to its expected
divergence at a quantum critical point. Above $T_{\mathbf{Q}}$, a
$\mathbf{Q}$-dependent scaling behavior of the form
$T\chi''(\mathbf{Q},E,T) =
C_{\mathbf{Q}}f[E/(a_{\mathbf{Q}}T^{\beta_{\mathbf{Q}}})]$ with
$\beta_{\mathbf{Q}}<1$ is obtained. This scaling does not enter
the general framework of quantum phase transition theories, since
it is obtained in a high temperature range, where Kondo spin
fluctuations depend strongly on temperature.

\end{abstract}


\maketitle

\section{Introduction}

In many heavy fermions systems (HFS) a quantum phase transition
(QPT) separates a non-magnetic from a magnetic ground state at
$T=0$ K. Such a transition is governed by the competition between
Kondo screening of localized moments and RKKY-like intersite
interactions. It can be tuned by applying an external pressure, a
magnetic field or by chemical substitution. In the vicinity of a
QPT, the critical fluctuations have a quantum feature
characterized by an effective dimension $d^{*}=d+z$, $d$ being the
spatial dimension and $z$ the dynamical
exponent\cite{Hertz,Millis,Continentino,Sachdev}. The extra
dimension $z$ is related to the imaginary time direction ($z=2$
for antiferromagnetic fluctuations). When $T$ is increased, a
cross-over regime (also called quantum classical (QC)) sets up
when the fluctuations lose their quantum features and become
controlled by $T$. The dimension is then reduced from $d^{*}$ to
$d$. In a simple picture this can be seen as a finite-size scaling
\cite{Plischke}, where the "finite size" $\tau_{T}\sim1/T$ of the
system in the time dimension is decreased when $T$ is increased
\cite{Sondhi} : the time dimension is then progressively
suppressed. If $\tau$ is the relaxation time of quantum
fluctuations, the quantum regime is the low $T$ regime for which
$\tau_{T}>\tau$, where the dynamical properties behave as
functions of $\omega\tau$ and do not depend on $T$. The cross-over
regime is then expected for $\tau_{T}<\tau$, its dynamical
properties behaving as functions of $\omega/T$.

Inelastic neutron
scattering (INS) is a unique tool for studying dynamic magnetic
properties. Enhancements of spin fluctuations (SF) have already
been reported by INS around the QPT of HFS (see for example
\cite{Raymond97,Stockert,Kadowaki1}). The $T$-dependence of those
low-energy excitations is believed to be related to the
low-temperature non Fermi liquid behavior observed by bulk
measurements near the quantum critical point (QCP) of such systems
\cite{Stewart}. That is why it is necessary for the understanding
of QPT to study precisely how SF evolve with $T$ and to search for
scaling laws specific to the QC regime. Several INS studies report
$\omega/T$ scaling of the dynamical spin susceptibility in HFS or
high $T_{C}$ superconductors
\cite{Schroder1,Aronson,Park,Montfrooj,Bao}. In particular, some
insight was given by the detailed study of Schr\"{o}der et al. at
the QCP of CeCu$_{6-x}$Au$_{x}$ \cite{Schroder1}: they obtained a
collapse of the dynamical spin susceptibility on a single curve
when plotted as $T^{0.75}\chi''(\omega,T)=g(\omega/T)$. A general
form of this law was found to work with the same $T$-exponents for
each vector of the reciprocal lattice and down to the smallest
accessible temperatures. This $\mathbf{Q}$-dependence became the
starting point of a local description of quantum criticality
\cite{Coleman,Si}. Such a description is opposed to itinerant
scenarii where the QPT is only driven by fluctuations at some
critical wavevectors \cite{Hertz,Millis,Moriya,Lavagna}.

We have chosen here to search for a scaling behavior at the QCP of
Ce$_{1-x}$La$_{x}$Ru$_{2}$Si$_{2}$, a HFS that has been
extensively studied for about 20 years
\cite{Fisher,Kambe,Regnault1,Raymond2001,Raymond2002}. This 3D
Ising system has a QCP at $x_{c}\simeq7.5\%$ that separates a
paramagnetic ground state for $x<x_{c}$ from an antiferromagnetic
ground state with the incommensurate propagation vector
$\mathbf{k_{1}}=$(0.31 0 0) for $x>x_{c}$. Although the occurrence
of small magnetic moments has been reported for $x\leq x_{c}$
(0.02 $\mu_{B}$ at $\mathbf{k_{1}}=$(0.31 0 0) below 2 K for
$x=x_{c}$ \cite{Raymond97} and 0.001 $\mu_{B}$ also below 2 K for
$x = 0$ \cite{Amato1}), a long range magnetic order with diverging
correlation length is only obtained for $x>x_{c}$. Large single
crystals are available, which makes it possible to investigate
precisely the reciprocal space via INS. In this system, the
observed excitation spectra consist in short range magnetic
correlations enhanced at the wavevectors
$\mathbf{k}_{1}$, $\mathbf{k}_{2}=(0.31,0.31,0)$, and
$\mathbf{k}_{3}=(0,0,0.35)$, while uncorrelated
SF are obtained away from these wavevectors and cover most of the
Brillouin zone (see Ref. \cite{Kadowaki2} for a detailed survey of
the SF repartition in the reciprocal space of this system).
Previous neutron measurements have shown the continuous behavior
of the SF through the QCP \cite{Raymond2001,Raymond2002} and
several tests led to a rather good accordance between Moriya's
itinerant theory and experimental data
\cite{Raymond97,Moriya,Kambe,Kadowaki2}.
Ce$_{1-x}$La$_{x}$Ru$_{2}$Si$_{2}$ constitutes consequently an
opportunity to study quantum criticality in a case for which the
itineracy of the $4f$ electrons is established. For this purpose,
we present here new measurements at the critical concentration
$x_{c}$ that were made not only to benefit from much better
statistics but also to measure a broader range of temperatures
(between 40 mK and 294 K) and energies (between 0.1 and 9.5 meV).
Such extended data are required for a precise determination of the
temperature dependence of the SF. In this paper we report an
anomalous scaling behavior of the dynamical spin susceptibility at
the QCP of Ce$_{1-x}$La$_{x}$Ru$_{2}$Si$_{2}$: instead of
$\omega/T$, $\omega/T^{\beta_{\mathbf{Q}}}$ scalings with
$\beta_{\mathbf{Q}}<1$ are obtained. Contrary to the other cases
reported in literature, the laws found here depend on the
wavevector, and each wavevector is characterized by a different
low-temperature cut-off below which a nearly $T$-independent
quantum regime is obtained.

\section{Experimental details}

The single crystals of Ce$_{0.925}$La$_{0.075}$Ru$_{2}$Si$_{2}$
studied here have been grown by the Czochralsky method. They
crystallize in the body centered tetragonal I4/mmm space group
with the lattice parameters $a = b = 4.197$ $\rm{\AA}$ and
$c=9.797$ \AA. A single crystal of 250 mm$^{3}$ was used for the
INS measurements and a smaller one of 3 mm$^{3}$ for the DC
susceptibility measurements. INS measurements were carried out on
the cold and thermal triple-axis spectrometers IN12 and IN22 at
the ILL (Grenoble, France). The $(001)$ plane was investigated.
60'-open-open and open-open-open set-up were used on IN12 and
IN22, respectively. A beryllium filter on IN12 and a pyrolytic
graphite (PG) filter on IN22 were added to eliminate higher-order
contaminations. In both cases PG was used for the vertically
focusing monochromator and for the horizontally focusing analyzer.
The final neutron energy was fixed to 4.65 meV on IN12 and to 14.7
meV on IN22 with the resulting energy resolutions of about 0.17
meV on IN12 and 1 meV on IN22 (FWHM of the incoherent signal). For
temperatures between 2.5 and 80 K the high-energy points obtained
on IN22 were combined with the ones obtained on IN12, with an
appropriate scale factor chosen for the collapse of the data in
their common range 1.9-2.5 meV. A complementary neutron experiment
was carried out on the inverted-geometry time-of-flight
spectrometer IRIS at ISIS (Didcot, U.K.) using a fixed final
neutron energy of 1.84 meV (PG analyzer) resulting in 18 $\mu$eV
resolution FWHM. The susceptibility measurements were performed
both in a commercial SQUID DC magnetometer for temperatures
between 5 and 300 K and in a dilution refrigerator SQUID DC
magnetometer for temperatures between 250 mK and 5 K, with the
magnetic field along the $[001]$ easy axis in both cases.

\section{Temperature dependence of spin fluctuations}

The data presented here consist in energy scans obtained by INS at
two wavevectors: the antiferromagnetic momentum transfer
$\mathbf{Q}_{1}$ = (0.69 1 0) = ${\bm\tau}$ - $\mathbf{k}_{1}$,
where ${\bm\tau}$ = (1 1 0) is a reciprocal lattice vector, and
the wavevector $\mathbf{Q}_{0}$ = (0.44 1 0), which is
sufficiently far from $\mathbf{k}_{1}$, $\mathbf{k}_{2}$, and
$\mathbf{k}_{3}$, so that no spatial correlations are observed. In
FIG. 1 the excitations spectra obtained at those two vectors are
plotted for three representative temperatures: their shape is
characteristic of a relaxation process. At $T=5$ K,
antiferromagnetic fluctuations are enhanced in comparison with the
ones obtained at $\mathbf{Q}_{0}$. When the temperature is raised
the difference between the two signals is attenuated and above the
correlation temperature $T_{corr}\simeq$ 80 K they are almost
identical; the system has lost its antiferromagnetic correlations.
We can also notice that the two signals are identical for $E>4$
meV at all temperatures. The observed intensity is proportional to
the scattering function $S(\mathbf{Q},E,T)$ (where $E=\hbar
\omega$), from which the imaginary part of the dynamical
susceptibility $\chi''(\mathbf{Q},E,T)$ is deduced using:
\begin{eqnarray}
  S(\mathbf{Q},E,T) &=&
  \frac{1}{\pi}\frac{1}{1-e^{-E/k_{B}T}}\chi''(\mathbf{Q},E,T).
\end{eqnarray}

\begin{figure}[h]
    \centering
    \epsfig{file=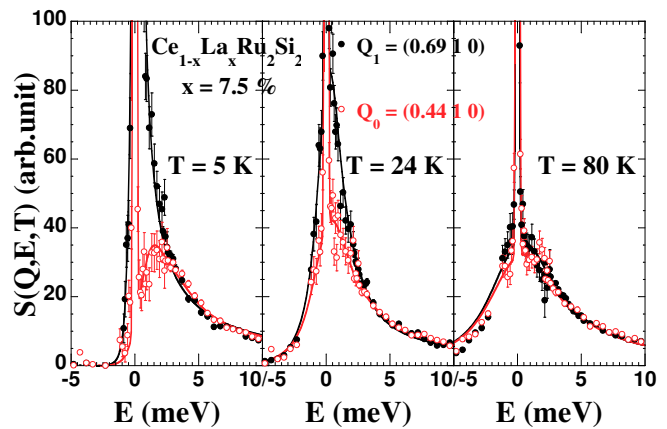,height=56mm}
    \caption{INS spectra obtained at $T=5$, 24 and 80 K for the momentum transfers $\mathbf{Q}_{1}$ and $\mathbf{Q}_{0}$.
    A constant background deduced from the scattering at low temperature and negative energy transfers has been subtracted.
    The scattering at $E=0$ corresponds to the incoherent elastic signal. The lines are fits to the data. (Color online) }
\end{figure}

For the two wavevectors the dynamical susceptibility is well
fitted by a single quasi elastic Lorentzian shape of the form
\cite{comment}:
\begin{eqnarray}
  \chi''(\mathbf{Q},E,T) &=& \frac{A(\mathbf{Q},T)}{\Gamma(\mathbf{Q},T)}\frac{E/\Gamma(\mathbf{Q},T)}{1+(E/\Gamma(\mathbf{Q},T))^{2}}
\end{eqnarray}
that corresponds to the simplest approximation that can be made to
treat the spin fluctuations. The general form of the dynamical
susceptibility is the Fourier transform of a single exponential
decay of relaxation rate $\Gamma(\mathbf{Q},T)$. It can be
expressed by:
\begin{eqnarray}
  \chi(\mathbf{Q},E,T) &=& \chi'(\mathbf{Q},E,T) + i \chi''(\mathbf{Q},E,T)\nonumber\\
  &= &\frac{A(\mathbf{Q},T)}{\Gamma(\mathbf{Q,T})-iE}.
\end{eqnarray}
In such a case, the static susceptibility is given by the
Kramers-Kronig relation:
\begin{eqnarray}
  \chi'(\mathbf{Q},T) = \chi'(\mathbf{Q},E=0,T) &=&\frac{1}{\pi}
  \int_{-\infty}^{\infty}\frac{\chi''(\mathbf{Q},E,T)}{E}dE \nonumber\\
  &= &\frac{A(\mathbf{Q},T)}{\Gamma(\mathbf{Q},T)}
\end{eqnarray}

However, in a previous thermal INS experiment on
CeRu$_{2}$Si$_{2}$, Adroja et al. observed a broad crystal field
(CF) excitation at about 30 meV that dominates the excitation
spectra for $E>$ 10 meV, its width (HWHM) being about 15 meV
\cite{Adroja}. Bulk susceptibility measurements also indicate that
the CF scheme do not change very much with concentration $x$ in
Ce$_{1-x}$La$_{x}$Ru$_{2}$Si$_{2}$ \cite{Fisher}. It is thus
reasonable to consider that in
Ce$_{0.925}$La$_{0.075}$Ru$_{2}$Si$_{2}$, as well as in
CeRu$_{2}$Si$_{2}$, the CF excitations dominate the low-energy SF
for $E>$ 10 meV. Instead of (4) it is finally better to
approximate the static susceptibility of the low-energy SF by
introducing an energy cut-off of 10 meV such as:
\begin{eqnarray}
  \chi'(\mathbf{Q},T) &=& \frac{2}{\pi}\int_{0}^{10}\frac{\chi''(\mathbf{Q},E,T)}{E}dE
\end{eqnarray}

\begin{figure}[h]
    \centering
    \epsfig{file=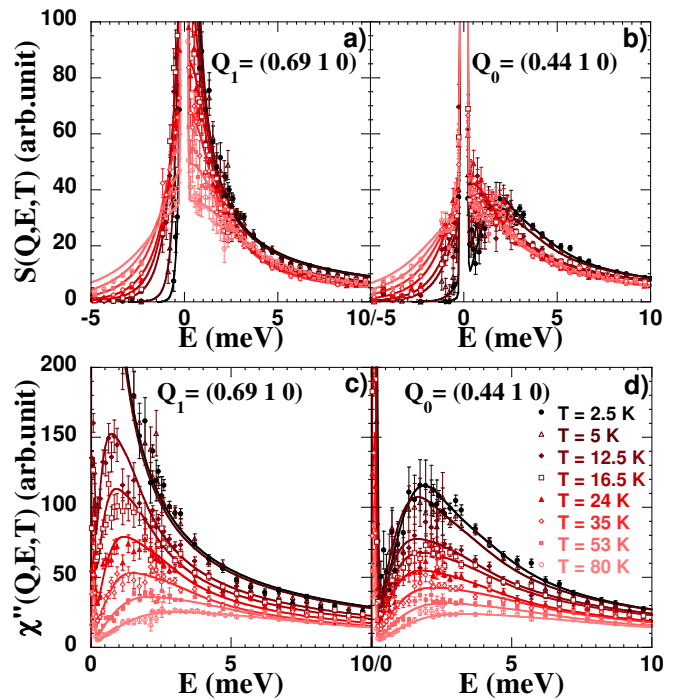,height=92mm}
    \caption{Scattering function $S(\mathbf{Q},E,T)$ a) at $\mathbf{Q}_{1}$ b) at
    $\mathbf{Q}_{0}$ and dynamical susceptibility $\chi''(\mathbf{Q},E,T)$ c) at $\mathbf{Q}_{1}$ d) at $\mathbf{Q}_{0}$ for $2.5<T<80$ K.
    The lines are fits to the data. (Color online)}
\end{figure}

$S(\mathbf{Q},E,T)$ and its fits using (2) are plotted for the two
wavevectors and 2.5 $<T<$ 80 K in FIG. 2 a and b.
$S(\mathbf{Q}_{1},E,T)$, which corresponds to antiferromagnetic
SF, is shown in FIG. 2 a: it is found to decrease in intensity and
to broaden when $T$ is increased. For uncorrelated SF, the
scattering intensity $S(\mathbf{Q}_{0},E,T)$, which is plotted in
FIG. 2 b, is characterized by the collapse of the data on a single
curve for positive energy transfers and $T>5$ K. The negative
energy points are strongly $T$-dependent because of the detailed
balance condition
$S(\mathbf{Q},-E,T)=exp(-E/k_{B}T)S(\mathbf{Q},E,T)$. Such a
behavior was also reported for the polycrystalline compounds
UCu$_{4}$Pd and CeRh$_{0.8}$Pd$_{2}$Sb, where the scattering is
temperature independent for positive energy transfers
\cite{Aronson,Park}. For $T=2.5$ and 5 K the uncorrelated signal
$S(\mathbf{Q}_{0},E,T)$ moves to higher energies. Although better
fits are obtained using an inelastic symmetrized Lorentzian
instead of the quasielastic Lorentzian shape (2), it is difficult
to conclude about their inelasticity, since the widths of these
peaks are too important. For both $\mathbf{Q}_{1}$ and
$\mathbf{Q}_{0}$, a strong $T$-dependence of the dynamical
susceptibility $\chi''(\mathbf{Q},E,T)$ deduced from (1) (and its
fits using (2)) is shown in FIG. 2 c and d. Contrary to
$S(\mathbf{Q},E,T)$, $\chi''(\mathbf{Q},E,T)$ has a decreasing
intensity for both wavevectors and is strongly broadened when $T$
is raised. Finally, for each spectrum, the relaxation rate
$\Gamma(\mathbf{Q},T)$ and the static susceptibility
$\chi'(\mathbf{Q},T)$ are extracted using (2) and (5). In the next
two subsections, the results of the fits of low-energy SF are
separately analyzed for the momentum transfers $\mathbf{Q}_{1}$
and $\mathbf{Q}_{0}$.

\subsection{Antiferromagnetic spin fluctuations}

\begin{figure}[h]
    \centering
    \epsfig{file=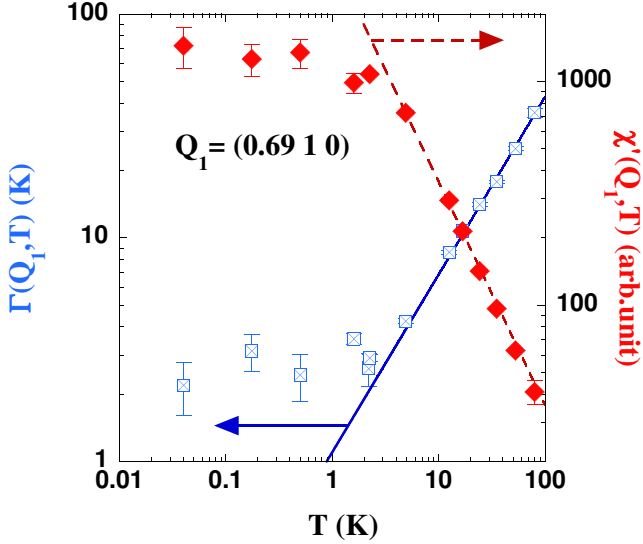,height=72mm}
    \caption{Temperature dependence of $\Gamma(\mathbf{Q}_{1},T)$ and $\chi'(\mathbf{Q}_{1},T)$.
    The full and dashed lines correspond to the high-temperature fits of the relaxation rate
    $\Gamma(\mathbf{Q}_{1},T)=1.1T^{0.8}$ and of the static susceptibility
    $\chi'(\mathbf{Q}_{1},T)=3550/T$, respectively. (Color online)}
    \end{figure}

The analysis of the antiferromagnetic SF at the momentum transfer
$\mathbf{Q}_{1}$ is only made below $T_{corr}\simeq$ 80 K. The
variations with $T$ of the relaxation rate
$\Gamma(\mathbf{Q}_{1},T)$ and the static susceptibility
$\chi'(\mathbf{Q}_{1},T)$ for antiferromagnetic SF are plotted in
FIG. 3. As seen, there are clearly two different regimes: a nearly
$T$-independent low-temperature and a strongly $T$-dependent
high-temperature regimes.

Below a characteristic temperature of $T_{1}\simeq3$ K,
$\chi''(\mathbf{Q}_{1},E,T)$ does not depend on $T$. Moreover, the
relaxation rate is found to have the value
$\Gamma(\mathbf{Q}_{1},T)\simeq k_{B}T_{1}$ in this regime: this
is thus the low-temperature regime for which $\tau<\tau_{T}$, with
$\tau=1/\Gamma(\mathbf{Q}_{1},T=0)\sim1/T_{1}$ and
$\tau_{T}\sim1/T$, as presented using a simple picture of scaling
in the Introduction. The saturation of antiferromagnetic SF
corresponds thus to their quantum regime. Because of the limited
resolution on IN12 and IN22, a complementary experiment was made
on the time-of-flight backscattering spectrometer IRIS.
Measurements were carried out at 100 mK and 2 K with a resolution
of 18 $\mu$eV. $\chi''(\mathbf{Q},E,T)$ was found to be
independent of $T$ for $\mathbf{Q}\simeq\mathbf{Q}_{1}$, which
confirms the saturation of antiferromagnetic SF at temperatures
below $T_{1}$.

At higher temperatures, $T_{1}<T<T_{corr}$, the antiferromagnetic
SF become controlled by $T$ such that $T$-power laws can be
extracted for $\chi'(\mathbf{Q}_{1},T)$ and
$\Gamma(\mathbf{Q}_{1},T)$:
\begin{eqnarray}
&\chi'(\mathbf{Q}_{1},T)  =& C_{1}/T^{\alpha_{1}}\;
  \;\;\; {\rm{and}} \;\;\;\; \Gamma(\mathbf{Q}_{1},T)  = a_{1}T^{\beta_{1}}
\end{eqnarray}
with
\begin{eqnarray}
\alpha_{1}=& 1\pm0.05 ,\;\;\;\;\;\;\;\;\;\;\;\;\;\;\;C_{1}  &= 3550\pm100 \;\;\rm{arb.\;unit,}\nonumber\\
\beta_{1} =& 0.8 \pm0.05, \;\;\; {\rm{and}}
\;\;\;\;a_{1}&=1.1\pm0.05\;\;\rm{SI\;unit.}\nonumber
\end{eqnarray}
To be more precise, the characteristic temperature $T_{1}$ has
been defined by the intercept of the two asymptotic regimes
obtained at low and high temperatures, the same intercept being
given for $\chi'(\mathbf{Q}_{1},T)$ and
$\Gamma(\mathbf{Q}_{1},T)$. Finally, the neutron data can be
plotted as $T\chi''(\mathbf{Q}_{1},E,T) =
C_{1}f[E/(a_{1}T^{0.8})]$ such that all the points measured for
$T_{1}<T<T_{corr}$ at the antiferromagnetic wavevector collapse on
the single curve $C_{1}f(x) =C_{1}x/(1+x^{2})$ with
$x=E/(a_{1}T^{0.8})$ (see FIG. 4). In the discussion, we will
focus on the anomalous form of this scaling law obtained for the
antiferromagnetic low-energy SF.

\begin{figure}[h]
    \centering
    \epsfig{file=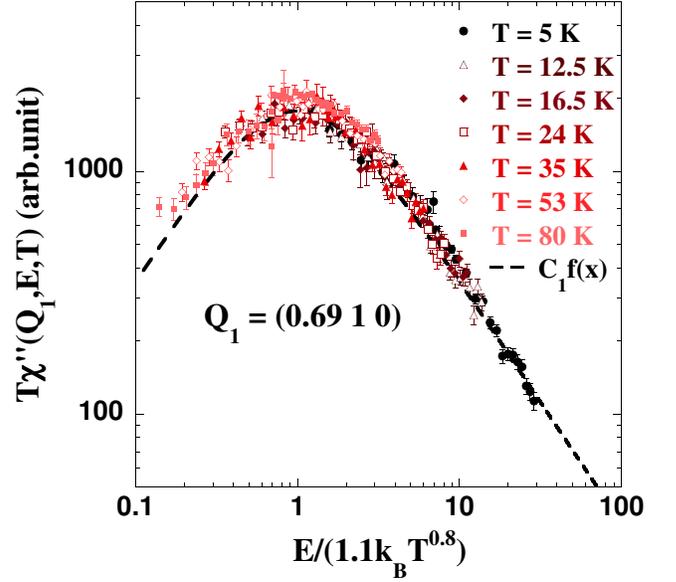,height=77mm}
    \caption{Scaling behavior of the low-energy antiferromagnetic SF obtained for $3<T<80$
K at $\mathbf{Q}_{1}$. The dynamical susceptibility follows the
scaling law $T\chi''(\mathbf{Q}_{1},E,T) =
C_{1}f[E/(a_{1}T^{0.8})]$ with $f(x)=x/(1+x^{2})$. (Color online)}
\end{figure}

\subsection{Uncorrelated spin fluctuations}

The temperature dependence of the relaxation rate
$\Gamma(\mathbf{Q}_{0},T)$ and the static susceptibility
$\chi'(\mathbf{Q}_{0},T)$ of uncorrelated SF are plotted in FIG.
5. As for the antiferromagnetic SF, we can define a characteristic
temperature $T_{0}\simeq 17$ K at the intercept of the asymptotic
low and high-temperature regimes. Below $T_{0}$,
$\chi''(\mathbf{Q}_{0},E,T)$ does not depend on $T$, and
$\Gamma(\mathbf{Q}_{0},T)\simeq k_{B}T_{0}$. For $T$ larger than
$T_{0}$, $T$-power laws can be extracted; the fits made on
$\chi'(\mathbf{Q}_{0},T)$ for $T\geq80$ K and on
$\Gamma(\mathbf{Q}_{0},T)$ for $T\geq20$ K give:
\begin{eqnarray}
\chi'(\mathbf{Q}_{0},T)  = C_{0}/T^{\alpha_{0}} \;\;{\rm{and}}
\;\;\Gamma(\mathbf{Q}_{0},T)  = a_{0}T^{\beta_{0}}
\end{eqnarray}
with
\begin{eqnarray}
\alpha_{0}=& 1\pm0.1 ,\;\;\;\;\;\;\;\;\;\;\;\;\;\;C_{0}  &= 2740\pm200 \;\;\rm{arb.\;unit,}\nonumber\\
\beta_{0} =& 0.6 \pm0.2, \;\;\; {\rm{and}}
\;\;\;\;a_{0}&=3.1\pm0.5\;\;\rm{SI\;unit.} \nonumber
\end{eqnarray}
\begin{figure}[h]
    \centering
    \epsfig{file=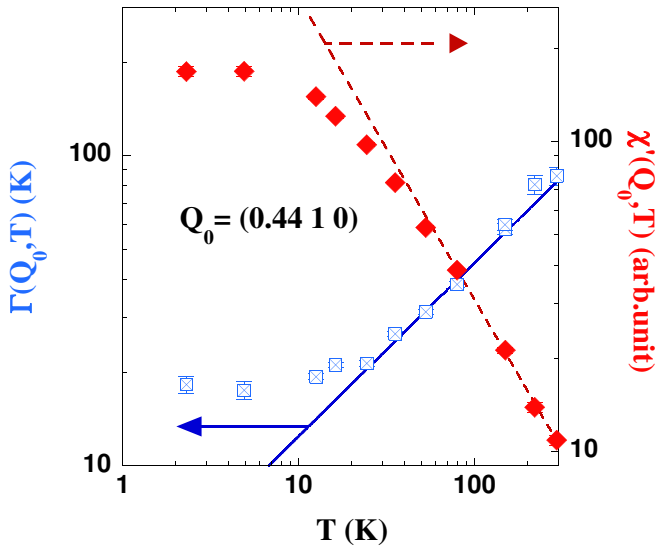,height=72mm}
    \caption{Temperature dependence of $\Gamma(\mathbf{Q}_{0},T)$ and $\chi'(\mathbf{Q}_{0},T)$. The full and dashed lines correspond
    to the high temperature fits of the relaxation rate $\Gamma(\mathbf{Q}_{0},T)=3.1T^{0.6}$ and of the static susceptibility
    $\chi'(\mathbf{Q}_{0},T)=2740/T$, respectively. (Color online)}
\end{figure}

However, higher energy scales and smaller intensities make the
study of uncorrelated SF more difficult than for the
antiferromagnetic case. Indeed, the corresponding quantum regime
differs from the antiferromagnetic one in its energy scale
$k_{B}T_{0}$ that is 5 times larger than the antiferromagnetic
energy scale $k_{B}T_{1}$. Then, the $T$-dependent regime must be
analyzed at temperatures sufficiently higher than $T_{0}$.
For those temperatures, the ground state SF and the CF excitations are no more completely separate entities in the magnetic excitation spectrum.
This affects the determination of $\alpha_{0}$ and
$\beta_{0}$ and makes their uncertainties larger than in the
antiferromagnetic case. Part of the uncertainty is removed by introducing a cut-off when determining the susceptibility with (5). This procedure is justified a posteriori by
the recovery of a Curie-like
behavior of $\chi'(\mathbf{Q}_{0},T)$ at high temperature in good agreement with
the bulk susceptibility (see section 4.1).
Since  the susceptibility
$\chi'(\mathbf{Q}_{0},T)$ estimated from (5) is notably different from the one estimated with
$A(\mathbf{Q}_{0},T)/\Gamma(\mathbf{Q}_{0},T)$,  a scaling plot for the uncorrelated fluctuations at $\mathbf{Q}_{0}$, as done in FIG. 4 for $\mathbf{Q}_{1}$, is not meaningful using the raw neutron data. However, the analysis of each individual spectrum that leads to (7) implies that $T\chi''(\mathbf{Q}_{0},E,T) =
C_{0}f[E/(a_{0}T^{\beta_{0}})]$.

\section{Discussion}

\subsection{Comparison with bulk susceptibility}

The bulk susceptibility $\chi_{bulk}$, measured along the $c$
axis, is compared in FIG. 6 with the microscopic static
susceptibilities $\chi'(\mathbf{Q}_{1},T)$ and
$\chi'(\mathbf{Q}_{0},T)$. For $T>100$ K, $\chi_{bulk}(T)$ follows
a Curie-Weiss law with a Curie temperature $\theta\simeq20$ K,
i.e. $\chi_{bulk}(T)=C/(T-\theta)$. For $T>100$ K, the static
susceptibilities $\chi'(\mathbf{Q},T)$ deduced from INS have also
been fitted by Curie-like laws that are undistinguishable from
Curie-Weiss laws, within the uncertainty in $\chi'(\mathbf{Q},T)$.
From the CF scheme of Ce$_{1-x}$La$_{x}$Ru$_{2}$Si$_{2}$
\cite{Lehmann, Lacerda}, it is known that the Van Vleck term is
negligible in the bulk $c$-axis susceptibility \cite{Vanvleck}.
Since the high-energy CF excitation is not taken into account in
the integrated susceptibilities from INS, both macroscopic and
microscopic susceptibilities $\chi_{bulk}(T)$ and
$\chi'(\mathbf{Q},T)$ correspond only to low-energy SF. The bulk
susceptibility being a measure at the wavevector $\mathbf{Q}=0$,
we have $\chi_{bulk}(T)=\chi'(\mathbf{Q}=0,T)$ when
$\chi'(\mathbf{Q},T)$ is obtained using (5). For $T>T_{corr}$,
there are no more magnetic correlations and $\chi'(\mathbf{Q},T)$
does not depend on $\mathbf{Q}$. It is thus adequate to adjust
$\chi'(\mathbf{Q},T)$ to  $\chi_{bulk}(T)$ at high temperatures,
as shown in FIG. 6.

As for $\chi'(\mathbf{Q}_{1},T)$ and $\chi'(\mathbf{Q}_{0},T)$, we
can define for $\chi_{bulk}(T)$ a characteristic temperature
$T^{*}\simeq16$ K at the intercept of the two asymptotic low and
high-temperature regimes. For $T<T_{corr}$, the fluctuations are
spatially-correlated, and the hierarchy
$\chi'(\mathbf{Q}_{0},0)<\chi_{bulk}(0)\ll\chi'(\mathbf{Q}_{1},0)$
is obtained in the low temperature quantum regime. This means that
low energy SF are slightly more important at $\mathbf{Q}=$ 0 than
at $\mathbf{Q}_{0}$, both being much smaller than the
antiferromagnetic SF. The slight enhancement of the SF at
$\mathbf{Q}=$ 0 is most likely due to weak ferromagnetic
correlations and is linked to the metamagnetic transition of the
system Ce$_{1-x}$La$_{x}$Ru$_{2}$Si$_{2}$: the application of a
magnetic field induces an increase of ferromagnetic SF that are
maximal at the metamagnetic field $H_{m}$
\cite{Raymond98,Fisher,Haen,Flouquet1,Flouquet2}.

\begin{figure}[h]
    \centering
    \epsfig{file=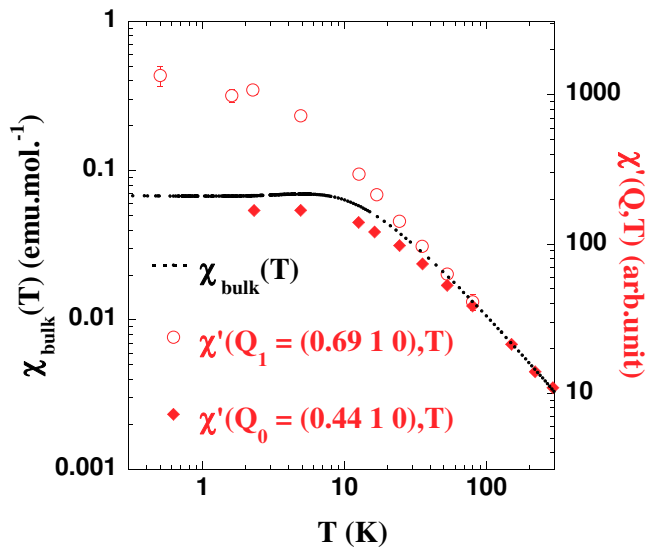,height=72mm}
    \caption{Temperature dependence of $\chi'(\mathbf{Q}_{1},T)$
    and $\chi'(\mathbf{Q}_{0},T)$ in comparison with $\chi_{bulk}(T)$.
    (Color online)}
\end{figure}

At low temperatures, we can also express $\chi'(\mathbf{Q}_{1},T)$
and $\chi'(\mathbf{Q}_{0},T)$ in CGS units, which gives:
\begin{eqnarray}
T_{1}.\chi'(\mathbf{Q}_{1},0)  = 1.00\pm0.1 \;\;\rm{K.emu.mol^{-1}
}, \nonumber \\
T_{0}.\chi'(\mathbf{Q}_{0},0)  = 0.97\pm0.1
\;\;\rm{K.emu.mol^{-1}}, \nonumber\\
and\;\;\;T^{*}.\chi_{bulk}(0)  = 1.09\pm0.1\;\; \rm{K.emu.mol^{-1}
.\nonumber}
\end{eqnarray}
If $T_{\mathbf{Q}}$ is the characteristic temperature of the SF at
the wavevector $\mathbf{Q}$, we have thus
$T_{\mathbf{Q}}.\chi'(\mathbf{Q},0)$ independent of $\mathbf{Q}$,
within the error bars. Since $\Gamma(\mathbf{Q}_{1},0)\simeq
k_{B}T_{1}$ and $\Gamma(\mathbf{Q}_{0},0)\simeq k_{B}T_{0}$, we
can assume $\Gamma(\mathbf{Q}=0,0)\simeq k_{B}T^{*}$, so that the
product $\Gamma(\mathbf{Q},0).\chi'(\mathbf{Q},0)$ is independent
of $\mathbf{Q}$. Hence, the low-temperature magnetic properties
are in good agreement with a Fermi liquid description of a
correlated system governed by an antiferromagnetic instability,
for which $\Gamma(\mathbf{Q},T).\chi'(\mathbf{Q},T)$ is expected
to be constant \cite{Moriya,Kuramoto1,Kuramoto2}. This Fermi
liquid picture is broken when, at high temperatures,
$\Gamma(\mathbf{Q},T).\chi'(\mathbf{Q},T)$ drops because of the
different $T$-behaviors of $\Gamma(\mathbf{Q},T)$ and
$1/\chi'(\mathbf{Q},T)$.

\subsection{Theoretical scenarii}

Contrary to the simple picture of scaling presented in the
Introduction \cite{Sondhi} and also to the different cases of
scaling reported in the literature
\cite{Aronson,Schroder1,Park,Montfrooj,Bao}, we obtain
$\omega/T^{\beta_{\mathbf{Q}}}$ instead of $\omega/T$ scalings of
the dynamical spin susceptibility of low-energy SF. We have showed
that the low-energy SF of Ce$_{0.925}$La$_{0.075}$Ru$_{2}$Si$_{2}$
obey scaling laws that depend on the wavevector $\mathbf{Q}$, each
one being characterized by a different low-temperature cut-off
$T_{\mathbf{Q}}$, below which a $T$-independent Fermi liquid-like
quantum regime is obtained. While a local description of quantum
criticality was proposed to explain the behavior of
CeCu$_{6-x}$Au$_{x}$ \cite{Coleman,Si,Schroder1}, the itinerant
character of our system is a key element to understand its
behavior \cite{Raymond97,Kambe,Raymond2001,Raymond2002}, and a
scenario for which the QPT is driven by itinerant magnetism should
be preferred. However, our system does not enter the framework of
existing itinerant theories for QPT
\cite{Hertz,Millis,Moriya,Lavagna}: two main discrepancies are
obtained between theoretical and experimental features.

A first disagreement comes from the saturation below a finite
temperature $T_{1}\simeq$ 3 K of $\chi''(\mathbf{Q}_{1},E,T)$: we
do not observe any divergence of the dynamical spin susceptibility
at the QPT of Ce$_{1-x}$La$_{x}$Ru$_{2}$Si$_{2}$ and a
low-temperature cut-off has to be taken into account. The
saturation of antiferromagnetic SF at the QPT of this system was
already reported for both cases of tuning by concentration or
pressure \cite{Raymond2001,Raymond2002}. The origin of this
cut-off is not yet well understood. It could be linked to the
appearance of a tiny magnetic moment below $T_{m}=2$ K $\simeq
T_{1}$ \cite{Raymond97}. The saturation of the dynamical spin
susceptibility is in marked contrast with its expected divergence
at a quantum critical point.

The second discrepancy comes from that, in the QC regime,
itinerant SF theories \cite{Hertz,Millis,Moriya,Lavagna} predict
for 3D antiferromagnetic SF a $\omega/T^{\beta}$ scaling law with
$\beta$ = 3/2 instead of our experimental $\beta_{1}$ = 0.8; more
generally, a value of $\beta$ smaller than 1 cannot be obtained in
the theories of QPT \cite{Continentino,Sachdev}. In SF theories
\cite{Moriya}, a mean-field picture is used to build a
$\mathbf{Q}$-dependent dynamical susceptibility
$\chi(\mathbf{Q},E,T)$ from a bare susceptibility $\chi_{0}(E,T)$
:
\begin{eqnarray}
 1/\chi(\mathbf{Q},E,T)=1/\chi_{0}(E,T)-J(\mathbf{Q},T)
\end{eqnarray}
where $J(\mathbf{Q},T)$ is the exchange interaction. In
Ce$_{0.925}$La$_{0.075}$Ru$_{2}$Si$_{2}$, the correlated signal
corresponds only to a small part of the Brillouin zone
\cite{Kadowaki2}. This fact together with the saturation of the
corresponding signal imply that the correlated response has a
small spectral weight (of about 10 $\%$) when integrating the
dynamical spin susceptibility over the Brillouin zone. The bare
susceptibility can thus be approximated by the susceptibility
measured at $\mathbf{Q}_{0}$. The principle mechanism of
relaxation of 4$f$ electrons contributing to the bare
susceptibility is the Kondo effect. The Kondo temperature,
$T_{K}$,  is then usually estimated by the low temperature neutron
linewidth.  This will lead here to $T_{K}=T_{0}\simeq17$ K. This
estimation of $T_{K}$ is in very good agreement with the values
deduced from thermodynamic and thermoelectric power measurements
\cite{Lehmann,Amato2} and the approximation of considering the
susceptibility at $\mathbf{Q}_{0}$ for the bare susceptibility is
thus reasonable. In SF theories, the bare susceptibility is
supposed to be weakly temperature dependent. This corresponds to a
low temperature regime below the Kondo temperature where the $4f$
moments on cerium sites are screened by conduction electrons and
where the renormalized Fermi surface is fully formed. On the
contrary, we experimentally found a scaling in a temperature range
where the bare susceptibility has a strong temperature dependence.
This is certainly the main reason why unexpected exponents are
found. Since antiferromagnetic SF saturate below $T_{1}\simeq 3$
K, a search for the $\beta=$ 3/2 scaling predicted by SF theories
can consequently only be done in the range $T_{1}<T<T_{0}$, which
is experimentally very difficult to verify because of the
closeness of $T_{1}$ and $T_{0}$. The antiferromagnetic SF being
built from the bare one, these two quantities are not independent.
A theory including the temperature variation of the bare Kondo
susceptibility is thus needed to explain the anomalous scaling law
we obtain for the susceptibility at the antiferromagnetic
wavevector $\mathbf{Q}_{1}$ for $T>T_{1}$. An impurity Kondo model
seems to be a good starting point to describe the bare
susceptibility measured at $\mathbf{Q}_{0}$, since it leads, for
$T$ sufficiently higher than $T_{K}(=T_{0})$, to the Curie-Weiss
static susceptibility and to the $T^{1/2}$-like behavior of the
relaxation rate \cite{Bickers,Kuroda,Lopes,Loidl,Hewson}. Indeed,
we experimentally obtained at high temperature for the wavevector
$\mathbf{Q}_{0}$ a Curie susceptibility and a value of $\beta_{0}$
quite close to 0.5.

\subsection{Comparison with other compounds.}

In the present study, we have determined the energy and
temperature dependence of the SF at the QCP of
Ce$_{1-x}$La$_{x}$Ru$_{2}$Si$_{2}$. The use of single crystals on
triple-axis spectrometers allowed us to obtain information on the
SF at two wavevectors $\mathbf{Q}$, where different scaling
behaviors have been obtained. In earlier works on the scaling
properties of the dynamical spin susceptibility near the QPT of
other HFS, such as UCu$_{5-x}$Pd$_{x}$,
Ce(Ru$_{1-x}$Fe$_{x}$)$_{2}$Ge$_{2}$, and CeRh$_{1-x}$Pd$_{x}$Sb
\cite{Aronson,Park,Montfrooj}, the use of polycrystalline samples
on time-of-flight spectrometers made it more difficult to
establish with precision any $\mathbf{Q}$-dependence of the SF.

\begin{table}
\caption{Comparison of characteristic physical quantities of
CeCu$_{6}$ and
CeRu$_{2}$Si$_{2}$\cite{Kambe,Amato3,Regnault2,Rossat,Wilhelm,
Raymond2001,Raymond2002}}
\begin{ruledtabular}
\begin{tabular}{lcr}
&CeCu$_{6}$&CeRu$_{2}$Si$_{2}$\\
\hline
$\gamma$\footnotemark[1]& 1.5 JK$^{-2}$mol$^{-1}$ & 360 mJK$^{-2}$mol$^{-1}$\\
$T^{*}_{\gamma}$\footnotemark[1]& 0.2 K & 3 K\\
$T^{*}_{\rho}$\footnotemark[2]& 0.1 K & 0.3 K\\
$T_{0}$& 5 K & 23 K\\
$T_{1}$& 2 K & 10 K\\
$T_{corr}$& 4 K & 40 K\\
$P_{c}$\footnotemark[3]& -4 kbar & -3 kbar\\
\end{tabular}
\end{ruledtabular}
\footnotetext[1]{$C(T)/T=\gamma$ for $T<T^{*}_{\gamma}$}
\footnotetext[2]{$\rho(T)=\rho(0)+AT^{2}$ for $T<T^{*}_{\rho}$}
\footnotetext[3]{Corresponding pressures of the QCP of
CeCu$_{6-x}$Au$_{x}$ and
Ce$_{1-x}$La$_{x}$Ru$_{2}$Si$_{2}$}\end{table}

In the case of the QCP of the HFS CeCu$_{6-x}$Au$_{x}$ (obtained
for $x_{c}=0.1$),  Schr\"{o}der et al. benefited from the use of a
single crystal and from the combination of triple-axis and
time-of-flight techniques \cite{Schroder1,Schroder2}. The first
study, using a triple-axis spectrometer, established a scaling law
of the form $T^{0.75}\chi''(\omega,T)=g(\omega/T)$ at an
antiferromagnetic wavevector \cite{Schroder2}. However, the
extension of this law to other parts of the reciprocal lattice was
done using a time-of-flight spectrometer \cite{Schroder1}, which
limits the information that can be obtained concerning the
$\mathbf{Q}$-dependence. Nevertheless, they found a single general
form of scaling for every $\mathbf{Q}$ of the reciprocal lattice.
They also found their scaling law to work down to $T=0$ K, as
theoretically expected for critical SF at a QCP. In TABLE I are
reported the main physical quantities that characterize the
paramagnetic heavy fermion compounds CeCu$_{6}$ and
CeRu$_{2}$Si$_{2}$ at low temperatures. For those two compounds, a
Fermi liquid regime is obtained at low temperatures and
characterizes their strong coupling renormalized state: the linear
coefficient $\gamma$ of the specific heat is found to be constant
and highly renormalized for temperatures $T<T^{*}_{\gamma}$
\cite{Kambe}, and the resistivity behaves as $\rho(0)+AT^{2}$ for
$T<T^{*}_{\rho}$ \cite{Kambe,Amato3}. The temperatures $T_{1}$,
$T_{0}$, and $T_{corr}$ have been obtained by INS
\cite{Regnault2,Rossat}, as in the present study. As seen in TABLE
I, the characteristic temperatures of CeCu$_{6}$ are about 5-10
times smaller than the corresponding ones of CeRu$_{2}$Si$_{2}$.
Moreover, the QCP of CeCu$_{6-x}$Au$_{x}$ and
Ce$_{1-x}$La$_{x}$Ru$_{2}$Si$_{2}$ are separated from their parent
compounds CeCu$_{6}$ and CeRu$_{2}$Si$_{2}$ by the respective
equivalent pressures of -4 kbar and -3 kbar \cite{Wilhelm,
Raymond2001,Raymond2002}. Because of these similar pressures, we
believe that for CeCu$_{5.9}$Au$_{0.1}$, the characteristic
temperatures are finally also about 5-10 times smaller than those
of Ce$_{0.925}$La$_{0.075}$Ru$_{2}$Si$_{2}$. As well as the
cut-off temperature $T_{1}\simeq3$ K is found to characterize the
critical SF at the QCP of Ce$_{1-x}$La$_{x}$Ru$_{2}$Si$_{2}$, the
QCP of CeCu$_{6-x}$Au$_{x}$ could thus have a cut-off temperature
for its critical SF of order 0.3-0.6 K. Because of smaller
characteristic temperatures and energies, in
CeCu$_{5.9}$Au$_{0.1}$ the quantum regime of the critical SF is
thus much more difficult to distinguish from the classical scaling
regime, and no saturation of SF at low temperatures has been yet
established by INS.

Finally, our results are quite similar to those of a recent work
made by Bao et al. \cite{Bao}. They measured by INS the
antiferromagnetic fluctuations of a single crystal of
La$_{2}$Cu$_{0.94}$Li$_{.06}$O$_{4}$, using a triple-axis
spectrometer. Contrary to the former systems, this system is not a
HFS and is located in the Fermi liquid ground state region in the
vicinity of a QCP. However, as in the present work, Bao et al.
obtained a low-temperature quantum regime for which the dynamical
susceptibility is found not to depend on $T$, and a
high-temperature regime for which a scaling behavior is obtained,
the relaxation rate $\Gamma(T)$ being in this case proportional to
$T$ and the static susceptibility $\chi(T)$, deduced from INS,
following a Curie law.

\section{Conclusion}

A detailed study of the $T$-dependence of SF in
Ce$_{0.925}$La$_{0.075}$Ru$_{2}$Si$_{2}$ has been carried out in
this work. For each of the two wavevectors $\mathbf{Q_{1}}$ and
$\mathbf{Q_{0}}$, which correspond to antiferromagnetically
correlated and to uncorrelated SF, respectively, a cut-off
temperature $T_{\mathbf{Q}}$ delimits a low-temperature
$T$-independent Fermi liquid-like quantum regime from a
high-temperature scaling regime governed by $T$. The cut-off
temperatures $T_{1}\simeq3$ K and $T_{0}\simeq17$ K are obtained
at $\mathbf{Q_{1}}$ and $\mathbf{Q_{0}}$, respectively. Several
discrepancies with itinerant theories of QCP have been
established: i) at low temperatures, while antiferromagnetic SF
are enhanced in comparison with the uncorrelated ones, they
saturate below $T_{1}$ and thus do not diverge when $T$ tends to
zero. ii) For each wavevector, high-temperature $T$-power laws can
be extracted for the static susceptibility and the relaxation
rate, so that the dynamical spin susceptibility is found to follow
an anomalous scaling of the form $T\chi''(\mathbf{Q},E,T) =
C_{\mathbf{Q}}f[E/(a_{\mathbf{Q}}T^{\beta_{\mathbf{Q}}})]$ above
$T_{\mathbf{Q}}$. Anomalous exponents $\beta_{\mathbf{Q}}<1$ are
observed, which is incompatible with QPT theories. This is
probably because these scaling laws are obtained in a $T$-range
where Kondo SF are temperature dependent. Even at the QCP of an
itinerant heavy fermion system, a Kondo impurity scaling should
thus be taken into account as a starting point to understand the
antiferromagnetic scaling.

\section*{Acknowledgments}

We thank D.T. Adroja and B.D. Rainford for sending us unpublished
results on CF measurements in CeRu$_{2}$Si$_{2}$, and M.A.
Continentino, M. Lavagna, C. P\'{e}pin, B. Coqblin, C. Lacroix, S.
Burdin, Y. Sidis, A. Murani, N. Bernhoeft, and L.P. Regnault for
very useful discussions.

\end{document}